\documentclass{jfm}
 
\usepackage{graphicx}
\usepackage{natbib}
\usepackage{comment}
\usepackage[outercaption]{sidecap} 
\usepackage{url}
\usepackage{xcolor}
\usepackage{xfrac}
\usepackage{gensymb}  
\usepackage{siunitx}

\usepackage{amsmath}
\usepackage{subcaption}
\usepackage{caption}

\newcommand{\sidecaption}[1]
{\raisebox{\abovecaptionskip}{\begin{subfigure}[t]{1.6em}
  \caption[singlelinecheck=off]{}
  \label{#1}
\end{subfigure}}\ignorespaces}

\usepackage{tipa}

\let\realverbatim=\verbatim
\let\realendverbatim=\endverbatim
\renewcommand\verbatim{\par\addvspace{6pt plus 2pt minus 1pt}\realverbatim}
\renewcommand\endverbatim{\realendverbatim\addvspace{6pt plus 2pt minus 1pt}}
\makeatletter
\newcommand\verbsize{\@setfontsize\verbsize{10}\@xiipt}
\renewcommand\verbatim@font{\verbsize\normalfont\ttfamily}
\makeatother

\ifCUPmtlplainloaded \else
  \checkfont{eurm10}
  \iffontfound
    \IfFileExists{upmath.sty}
      {\typeout{^^JFound AMS Euler Roman fonts on the system,
                   using the 'upmath' package.^^J}%
       \usepackage{upmath}}
      {\typeout{^^JFound AMS Euler Roman fonts on the system, but you
                   don't seem to have the}%
       \typeout{'upmath' package installed. jfm.cls can take advantage
                 of these fonts,^^Jif you use 'upmath' package.^^J}%
      }
  \else
  \fi
\fi

\ifCUPmtlplainloaded \else
  \checkfont{msam10}
  \iffontfound
    \IfFileExists{amssymb.sty}
      {\typeout{^^JFound AMS Symbol fonts on the system, using the
                'amssymb' package.^^J}%
       \usepackage{amssymb}%
         \let\leq=\leqslant
         
      }{}
  \fi
\fi

\ifCUPmtlplainloaded \else
  \IfFileExists{amsbsy.sty}
    {\typeout{^^JFound the 'amsbsy' package on the system, using it.^^J}%
     \usepackage{amsbsy}}
    {\providecommand\boldsymbol[1]{\mbox{\boldmath $##1$}}}
\fi

\sbox{\astrutbox}{\rule[-5pt]{0pt}{20pt}}

\title[Journal of Fluid Mechanics]{\LaTeXe\ Input Guide for Authors}

\author{Cheng-Nian Xiao \and Inanc Senocak\corresp{\email{senocak@pitt.edu}}}
\affiliation{Department of Mechanical Engineering and Materials Science, \\University
of Pittsburgh, Pittsburgh, PA, USA 15261 }
\date{10 August 2001 and in revised form 10 August 2004}
\pubyear{2001} 
\volume{000}
\pagerange{\pageref{firstpage}--\pageref{lastpage}}
\doi{S002211200100456X}

\begin{document}

 
\title{Linear Stability of Katabatic Slope Flows with Ambient Wind Forcing}
\maketitle
\begin{abstract}
We investigate the stability of katabatic slope flows over an infinitely wide and uniformly cooled planar surface subject to an additional forcing due to a uniform downslope  wind field aloft. We adopt an extension of Prandtl's original model for slope flows \citep{lykosov1972} to derive the base flow, which constitutes an interesting basic state in stability analysis because it cannot be reduced to a single universal form independent of external parameters.  
We apply a linear modal analysis to this  basic state to demonstrate that for a fixed Prandtl number and  slope angle, two independent dimensionless parameters are sufficient to describe the flow stability. One of these parameters is the \textit{stratification perturbation number} that we have introduced  in \cite{xiao2019}. The second  parameter, which we will henceforth designate the \textit{wind forcing  number}, is hitherto uncharted and can be interpreted as the ratio of the kinetic energy of the ambient wind aloft to the  damping due to viscosity and stabilizing effect of the background stratification. 
For a fixed Prandtl number, stationary transverse and travelling longitudinal modes of instabilities can emerge, depending on the value of the slope angle and the aforementioned dimensionless numbers.
The influence of ambient wind forcing on the base flow's stability is complicated as the ambient wind can be both stabilizing as well as destabilizing for a certain range of the parameters. Our results constitute a strong counter-evidence against the current practice of relying solely on the gradient Richardson number to describe the dynamic stability of stratified atmospheric slope flows.
\end{abstract}
\vspace{-5mm}
 \section{Introduction}
 Ludwig Prandtl's slope flow model permits an exact solution to the Navier-Stokes equations including heat transfer at an infinitely wide inclined surface immersed within a stably stratified medium \citep{prandtl1942}. The model has been found to describe well the vertical profiles of wind speed and temperature associated with katabatic winds in mountainous terrain or over large ice sheets in (Ant-)arctica or Greenland \citep{fedorovich2009}. 

\begin{figure}
	\centering
	\includegraphics[width=0.3\textwidth]{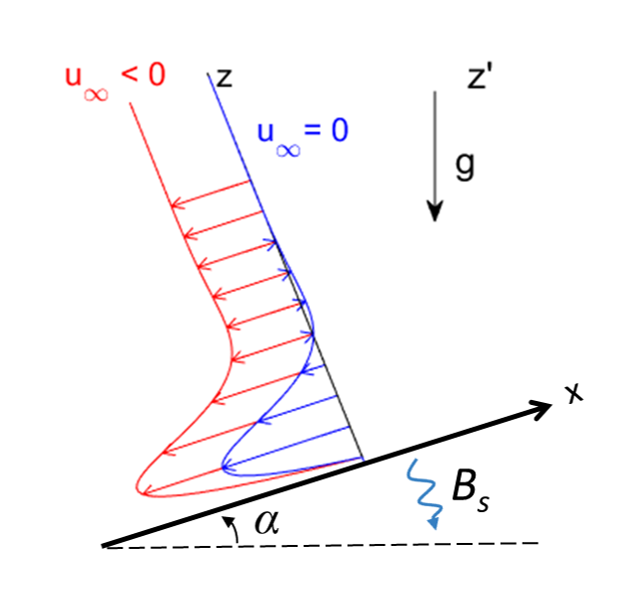}
	\vspace{-4mm}
	\caption{Velocity profiles corresponding to the extended  (red) and the original Prandtl model (blue) for slope flows. A rotated coordinate system is adopted.}
	\label{fig:sketchslope}
	\vspace{-4mm}
\end{figure}

Prandtl assumed quiescent winds at high altitudes in his slope model. It is also common for katabatic winds to develop in the presence of an external ambient wind field aloft, as for example when stably stratified air flows over a long mountain range, resulting in a non-zero wind in the free stream \citep{ manins1979field}. \citeauthor{lykosov1972} incorporated the effect of a uniform ambient wind field into Prandtl's original formulation.  We will henceforth refer to their model as the \textit{extended} Prandtl model. Katabatic wind profiles above an inclined cooled slope depicted by the original  and the extended  Prandtl model are shown in figure \ref{fig:sketchslope}. The vertical profiles of buoyancy and velocity as predicted by the original Prandtl model are exponentially damped sinusoidal solutions. In the original Prandtl model, the low-level jet along the slope descent is capped by a weak reverse flow. The extended Prandtl model appears as a mere shifting of the velocity profile produced by the original Prandtl model,however, as the equations indicate, the downslope ambient wind also increases  the velocity maximum of the low-level jet. This extended model can be accepted as a valid approximation to a situation in which stably stratified air flows over the top of an elevated terrain and follows the underlying surface closely \citep{whiteman2000}. In the present work, we adopt the extended Prandtl model with the assumption that the ambient wind is directed down-slope without cross-slope components and remains parallel to the inclined plane underneath.
  
  In \citet{xiao2019}, we  investigated the linear stability of the katabatic flows under the original Prandtl model and uncovered transverse and longitudinal modes of flow instabilities that emerge as a function of the slope angle, Prandtl number and a new dimensionless number, which we have designated the \textit{stratification perturbation} parameter.
  This new dimensionless number represent the importance of heat exchange at the surface relative to the strength of the ambient stratification, and it is defined solely by the intrinsic parameters of the flow problem at hand, and thus physically more meaningful than the more familiar \textit{internal Froude} or \textit{Richardson} numbers. However, by using ``derived'' internal length and velocity scales in the original Prandtl model, $\Pi_s$ can be converted a bulk Richardson or internal Froude number, creating a misleading interpretation that there is no necessity for such a new dimensionless parameter. In the present work, we demonstrate that $\Pi_s$ intrinsically exists along side with another new dimensionless number, and for fixed Prandtl number, these two dimensionless numbers along with the inclination angle describe the dynamic stability of stably stratified slope flows under the combined action of ambient wind and surface cooling. 
  Here, we pursue the same technical approach and the methods outlined in \cite{xiao2019} to determine the stability limits of the extended Prandtl model to comprehend the effect of a uniform ambient wind field on the stability katabatic slope flows.
\vspace{-6mm}
\section{Governing Equations}
Let us consider the slope flow under the action of an ambient wind as depicted in figure in figure~\ref{fig:sketchslope}, where
 $\alpha$  is the slope angle and $B_S$ is the constant negative heat flux imposed at the surface. The constant ambient wind speed in the free stream is $u_{\infty}$. For ease of analysis, the problem is studied in a rotated Cartesian coordinate system whose $x$ axis is aligned with the planar inclined surface and points along the upslope direction. 
 
 Let $b$ the scalar buoyancy variable, and $u$ be the along-slope (longitudinal), $v$ be the cross-slope (transverse), and $w$ be the slope-normal velocity components, such that $u_i=[u, v, w]$ is the velocity vector, where a positive value of $u$ is associated with the upslope direction. The gravity vector in the rotated coordinate system is then given by  $g_i=[g_1,g_2,g_3]=[\sin\alpha, 0, \cos\alpha]$, and we will also refer the spatial coordinate components $(x_j)$ in the rotated frame   as $(x,y,z)$. The ambient wind vector is assumed to be of the form $(U_j)=(u_{\infty},0,0) ,u_{\infty}<0$. The governing equations for conservation of momentum and energy under the Boussinesq approximation for an incompressible flow can be written as follows:
\begingroup
\allowdisplaybreaks
\begin{align}
\frac{\partial u_i}{\partial t}+\frac{\partial u_i u_j}{\partial x_j}
=&~ -\frac{1}{\rho}\frac{\partial p}{\partial x_i} + 
\frac{\partial}{\partial x_j} \left( \nu\frac{\partial u_i}{\partial x_j}\right) + bg_{i}, \label{eqnslopemom} \\
\frac{\partial b}{\partial t} +  \frac{ \partial b u_j}{\partial x_j}  = &~   \frac{\partial}{\partial x_j}\left( \beta\frac{\partial b}{\partial x_j} \right)  - N^2 g_j (u_j-U_j) \label{eqnslopebuoy},
\end{align}
\endgroup
where $\nu$ is the kinematic viscosity, $\beta$ is the thermal diffusivity. $N=\sqrt{\frac{g}{\Theta_r}\frac{\partial \Theta_e}{\partial z'}}$ is the Brunt-V\"ais\"al\"a frequency, assumed to be constant, $\Theta$ is the potential temperature, and $z'$ is the vertical coordinate in the non-rotated coordinate system.
Buoyancy is related to the potential temperature $\Theta$ as $b=g (\Theta-\Theta_e)/ \Theta_r$, where $\Theta_{r}$ is a reference potential temperature and $\Theta_e$ is the environmental potential temperature. The governing equations are completed by the divergence free velocity field condition for incompressible flows.


Following the same assumptions in the original Prandtl model, equations \ref{eqnslopemom}-\ref{eqnslopebuoy} reduce to simple momentum and buoyancy balance equations.
\cite{lykosov1972} presented an exact solution for the case with constant temperature at the surface and  ambient wind parallel to the surface. Here, we follow the approach presented in \cite{shapiro2004} and modify that solution for constant surface buoyancy flux at the surface. The modified solution takes the following form:
\begin{align}
u &=  [(u_{\infty}+\sqrt{2}u_0)\sin(z_n /\sqrt{2})  -u_{\infty}\cos(z_n /\sqrt{2})] \exp(-z_n/\sqrt{2}) +u_{\infty},\label{eqnprandtlsolutionu}\\ 
b &= \frac{2\nu}{z_0^2\sin \alpha }[((u_{\infty}+\sqrt{2}u_0)\cos(z_n/\sqrt{2})+u_{\infty}\sin(z_n/\sqrt{2}) ] \exp(-z_n/\sqrt{2}) ,\label{eqnprandtlsolutionb}
\end{align}
where $z_n=z/l_0$ is the nondimensional height, and the corresponding scales governing the flow problem are given as 
\begin{align}
l_0 = &~ (\nu\beta) ^{\sfrac{1}{4}} N^{\sfrac{-1}{2}}\sin^{\sfrac{-1}{2}}\alpha, \label{eqnzscale}\\
u_0 = &~ (\nu\beta) ^{\sfrac{-1}{4}}  N^{\sfrac{-3}{2}}B_{s} \sin^{\sfrac{-1}{2}}\alpha, \label{eqnuscale}\\
u_{c} = & u_0+u_{\infty}\label{eq:vc_scale}\\
b_0 = &~\left.\frac{\partial b}{\partial z}\right|_0 \cdot l_0= \frac{B_s}{\beta}l_0, \label{eqnbscale}
\end{align}
where $Pr\equiv{\nu}/{\beta}$ is the Prandtl number. A composite velocity scale $u_c$ is defined in equation \ref{eq:vc_scale} as the  sum of an inner velocity scale $u_0$ and an outer velocity scale that is the ambient wind speed $u_{\infty}$. It can be shown via calculus that for all values of $u_0,u_{\infty}<0$, the normalized maximal velocity $u_{\text{max}}/u_c$ of the flow profile as well as normalized location $z_{\text{max}}/l_0$ where this maximum is attained  always lie within a constant, finite interval independent of external flow parameters. 
Thus the choice of the velocity scale $u_c$ and length scale $l_0$ is both simple and meaningful for this class of flow profiles. We observe from (\ref{eqnprandtlsolutionu}) that the velocity profile exhibits the expected low-level jet near the surface and approaches the ambient wind speed $u_{\infty}$ at higher altitudes. This trend implies the existence of two distinct velocity scales, one that is associated with the processes near the surface based on the low-level jet, and another one that represents the ambient wind aloft. Thus, no matter which 
velocity scale is chosen, the flow profiles in the extended Prandtl model cannot be normalized to a universal form independent of the wind speed $u_{\infty}$, in contrast to the original Prandtl model with $u_{\infty}=0$.

%

Let us now consider the Buckingham-$\pi$ theorem to determine the dimensionless numbers involved in the extended Prandtl model for slope flows. One can show that any nondimensional dependent variable (e.g. nondimensional maximum jet velocity) is a function of the following four independent dimensionless parameters:
\begin{align}\label{eq:pi_set}
    \alpha, \quad Pr \equiv \frac{\nu}{\beta}, \quad \Pi_s \equiv \frac{|B_s|}{\beta N^2}, \quad \Pi_w \equiv \frac{u_{\infty}^2}{\nu N}
\end{align}
Due to the lack of an externally imposed length scale, familiar dimensionless numbers such as the Reynolds, Richardson, or Froude number do not appear in the above list and all the dimensionless numbers are functions of the externally imposed dimensional parameters in the slope flow problem only.

The new dimensionless number in the above set is $\Pi_w$. $\Pi_s$ was introduced in \cite{xiao2019}. We designate $\Pi_w$ the \textit{ wind forcing } number interpret it as the ratio of the kinetic energy in the ambient wind to the kinetic energy damping in the flow due to viscosity and stabilizing effect of stratification. 
\vspace{-4mm}
\section{Linear Stability Analysis}
We  introduce the normalized  velocity  and buoyancy as $u_n =u/u_c,b_n=b/b_0$, and use $l_0$ to normalize all lengths.
Linearizing around the base flow given by (\ref{eqnprandtlsolutionu})-(\ref{eqnprandtlsolutionb}),
and assuming that disturbances are waves of  the form $ \mathbf{q}(x,y,z,t) =\mathbf{\hat{q}}(z)\exp{\left \{ i(k_x x+k_y y)+ \textrm{\textomega} t \right \}} $, the resulting equations have the form
\begingroup
\allowdisplaybreaks
\begin{align}
ik_x\hat{u}+ik_y \hat{v}+ \frac{\partial \hat{w}}{\partial z} &= 0,\label{eqnslopelincont}\\
\textrm{\textomega} \hat{u}+iu_nk_x\hat{u}+u_n'\hat{w}
&= -ik_x\hat{p}-\frac{Pr \sin\alpha}{C}\left(-(k_x^2+k_y^2)\hat{u}+\frac{\partial^2\hat{u}}{\partial z^2}\right) -\frac{\Pi_s Pr\sin\alpha }{ {C}^2}\ \hat{b},\\
\textrm{\textomega} \hat{v}+iu_nk_x\hat{v}
&= -ik_y\hat{p}- \frac{Pr \sin \alpha}{C} \left(-(k_x^2+k_y^2)\hat{v}+\frac{\partial^2\hat{v}}{\partial z^2} \right),\\
\textrm{\textomega} \hat{w}+iu_nk_x\hat{w}
&= -\frac{\partial\hat{p}}{\partial z}-\frac{Pr \sin \alpha }{C}\left(-(k_x^2+k_y^2)\hat{w}+\frac{\partial^2\hat{w}}{\partial z^2} \right)
-\frac{\Pi_s Pr \cos \alpha }{{C} ^2} \ \hat{b} , \\
\textrm{\textomega} \hat{b}+iu_nk_x\hat{b}+b_n'\hat{w}
&=  -\frac{\sin\alpha }{C}\left(-(k_x^2+k_y^2)\hat{b}+\frac{\partial^2\hat{b}}{\partial z^2} \right)+ \frac{1}{\Pi_s}(  \hat{u} \sin\alpha+\hat{w} \cos\alpha)  ,\label{eqnslopelinlast}
\end{align}
\endgroup
where $\hat{u},\hat{v},\hat{w},\hat{p},\hat{b}$ are  flow disturbances varying along the slope normal direction normalised by $u_{c},b_{0}$, respectively. $z$ is the distance to the slope surface normalized by the length scale $l_0$. $k_x,k_y$ are normalized positive wavenumbers in the $x$ (along-slope) and $y$ (transverse) directions, respectively, whereas $\textrm{\textomega}$ is a normalized complex frequency. The normalised  base flow solution and its  derivative in the slope normal direction in normalised coordinates are denoted by $u_n,b_n$ and $u_n',b_n'$, respectively. The coefficient $C=\Pi_s + \sqrt{\Pi_w} Pr^{\sfrac{3}{4}}\sin^{\sfrac{1}{2}}\alpha$ is introduced solely for convenience, and we choose  $\Pi_w$ instead of $\sqrt{\Pi_w} Pr^{\sfrac{3}{4}}\sin^{\sfrac{1}{2}}\alpha$ as the  dimensionless parameter such that it is independent of the slope angle $\alpha$ and Prandtl number $Pr$ which are separate dimensionless numbers of the configuration.

The solution method for the above generalised eigenvalue problem follows the same approach as described in \cite{xiao2019}.
The stability behaviour of the problem is encoded by  the eigenvalues $\textomega$, whose real part equals the exponential growth rate and whose imaginary part is the temporal oscillation frequency of the corresponding eigenmode.
\begin{figure}
\centering
\sidecaption{subfig:kyRe30}\raisebox{-0.9\height}{
    \includegraphics[width=0.45\textwidth]{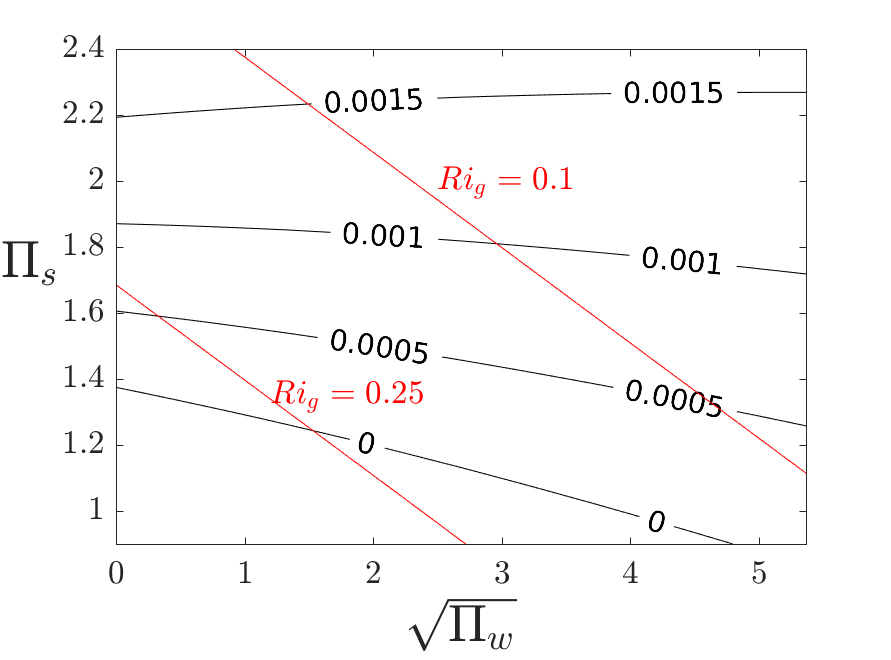}}~
\sidecaption{subfig:kxRe30}\raisebox{-0.88\height}{
    \includegraphics[width=0.44\textwidth]{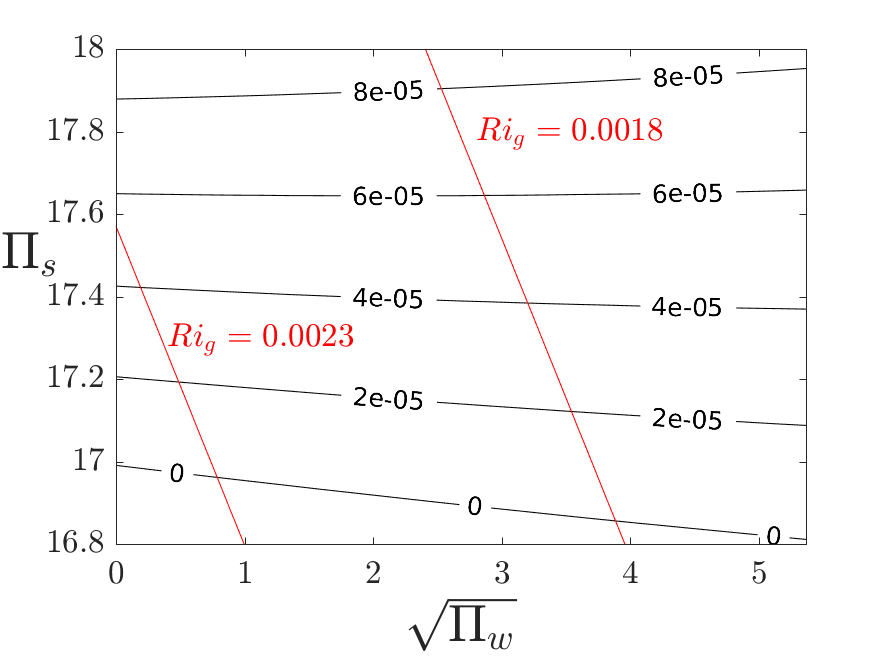}}
\sidecaption{subfig:kyRe66}\raisebox{-0.9\height}{
    \includegraphics[width=0.44\textwidth]{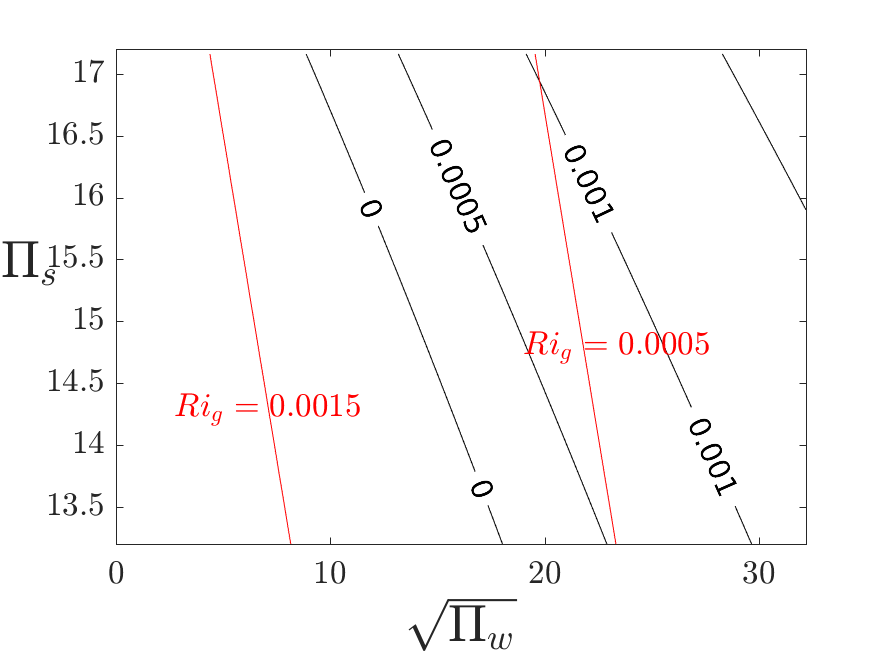}}~
\sidecaption{subfig:kxRe66}\raisebox{-0.9\height}{
    \includegraphics[width=0.45\textwidth]{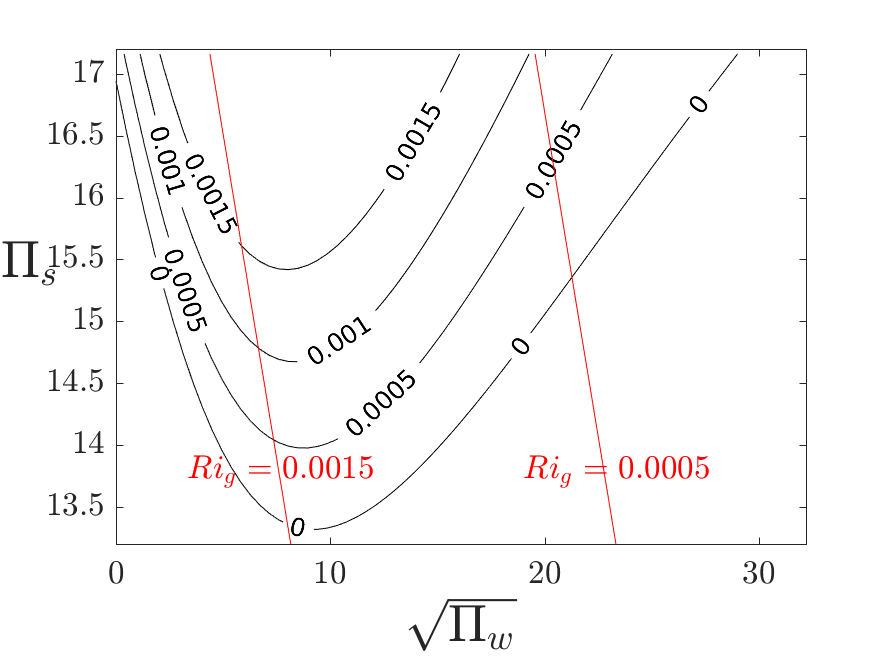}}
\caption{ Contours of maximal growth rates at $\alpha=4^{\circ}$ for (a) transverse and (b) longitudinal modes  depending on wind forcing number $\Pi_w$ and stratification perturbation $\Pi_s$.  The same contours at $\alpha=67^{\circ}$ for (c) transverse  and (d) longitudinal modes.  Red lines are contours
of the gradient Richardson number calculated from equation \ref{eq:Ri}.}
\label{fig:grcontourscrRe}
\vspace{-5mm}
\end{figure}

\subsection{Dependence of instability modes on dimensionless parameters}
From the results in \cite{xiao2019}, it is known that without ambient winds, the dominant instability of Prandtl's profile for katabatic flows  at each angle is either a stationary  transverse mode, i.e. varying purely along the cross-slope direction, or a longitudinal mode travelling  along-slope. This means that the instability growth rate as a function of the wave vectors $k_x,k_y$ attains its maximum on either the $k_x$ or the $k_y$ axis, with the other wave vector being zero. It turns out that the same also holds true for katabatic flows in the presence of ambient winds, hence the growth rate contours for  disturbances in the wave vector space, looking qualitatively similar like those in \cite{xiao2019}, will not be shown here. 

The fact that the most dominant instability  is either a  pure transverse  ($k_x=0$) or longitudinal mode ($k_y=0$)  means that at a fixed configuration determined by the slope angle $\alpha$ and the parameters $\Pi_s,\Pi_w$, the most dominant instability can be found  by searching for one of the wave numbers $k_x$ (longitudinal mode) or $k_y$(transverse mode) which maximizes the growth rate, setting the other wave number to zero. This approach has been applied to obtain the growth rate contour of the strongest transverse and longitudinal modes over the $\Pi_w,\Pi_s$ space for different slope angles, as shown in figure \ref{fig:grcontourscrRe}. At each given angle $\alpha$ and  ambient wind value specified by $\Pi_w$, we identify the most dominant amongst the transverse and longitudinal modes as the one which has a smaller $\Pi_s$ value for critical stability threshold, i.e. at which the growth rate is zero. 
Since the gradient Richardson number $Ri_g$  features prominently in the study of stratified flows, we overlay the corresponding $Ri_g$ number on the contour plots in figure \ref{fig:grcontourscrRe}. 
The $Ri_g$ number used in those figures is calculated from the extended Prandtl model velocity profile \ref{eqnprandtlsolutionu} and reduces to the following convenient formula with the help of the dimensionless parameters $\Pi_s,\Pi_w$ defined in equation \ref{eq:pi_set}
 \begin{equation}\label{eq:Ri}
     {Ri}_g =\frac{N^{2}}{\left( \frac{\partial u}{\partial z}\right)_{\text{max}}^{2}}
     = \frac{\text{Pr}}{\left(\sqrt{2\Pi_w}Pr^{3/4}\sin^{1/2}\alpha +\Pi_s \right)^2},
 \end{equation}
 where it can be shown that the maximum shear is attained at the slope surface.
 We can observe that $Ri_g$ number decreases with an increase of either of  the parameters $\Pi_s$ or $\Pi_w$. 
 However, since $\text{Ri}_g$ is a function depending on three independent variables $\alpha,\Pi_s, \Pi_w$ for a fixed $Pr$ number, it is possible from equation \ref{eq:Ri} to find different combinations of their values which lead to the same $Ri_g$ number.
 By inspecting the normalized partial derivative  $(\partial Ri_g/\partial \alpha)/Ri_g$, it can   be concluded that for   fixed $\Pi_s,\Pi_w\gg 1$, $Ri_g$ becomes insensitive  to variations of slope angle $\alpha$.
 This means that at those more unstable flow configurations, the $Ri_g$ number remains almost constant for all angles $\alpha$.
 But as the results shown in figure \ref{fig:grcontourscrRe} demonstrate,  different values for either of these dimensionless parameters can have profoundly different effects on the linear stability of the underlying base flow. For example, at different slope angles, the dominant instability may change from either the stationary transverse mode  or the travelling longitudinal mode to the other instability, respectively. From the plots shown in figure \ref{fig:grcontourscrRe}, we observe that increasing the ambient
 wind tend to lower an instability's growth rate at the same $Ri_g$ number, i.e. 
 the most unstable mode at fixed $Ri_g$ is found at $u_{\infty}=0$. which is the original Prandtl model as analysed in \cite{xiao2019}. At the low slope angle of $\alpha=4^{\circ}$, it  can be observed that  for the  wind forcing number $\Pi_w<3$, the base flow can be unstable despite possessing a larger $Ri_g$ number than critical value $Ri_g=0.25$. This counter example to the celebrated Miles-Howard stability theorem has already been shown in \cite{xiao2019} for the Prandtl base flow without ambient wind and has been attributed to the presence of surface  inclination, heat transfer at the surface, and viscosity.
 
We observe from figure \ref{fig:grcontourscrRe} that an increase of surface buoyancy, as measured by the dimensionless number $\Pi_s$, is a monotonically destabilizing effect for both the transverse and longitudinal modes. This observation is in complete agreement with the stability results for the original Prandtl model as demonstrated in \cite{xiao2019}. The effect of $\Pi_w$ on the instabilities, however, is slightly more complex.
Figures \ref{fig:grcontourscrRe}a and \ref{fig:grcontourscrRe}c indicate that for both slope angles $\alpha=4^{\circ},67^{\circ} $, the growth rate of the most unstable transverse mode grows monotonically with an increase in $\Pi_w$.  As shown in \cite{xiao2019},  at low slope angles devoid of an external ambient wind forcing (i.e. $u_{\infty}=0$), the transverse mode is the dominant instability. Thus at those angles,  when all other flow parameters are left unchanged, ambient wind has a strictly destabilizing effect on the base flow field. This behaviour is consistent with  expectation since increasing the ambient wind also increases the maximal shear of the base flow profile given in equation \ref{eqnprandtlsolutionb}, thus decreasing the $Ri_g$ number, according to equation \ref{eq:Ri}. 
\begin{figure}
\sidecaption{subfig:modes67}\raisebox{-0.9\height}{
\includegraphics[width=0.47\textwidth]{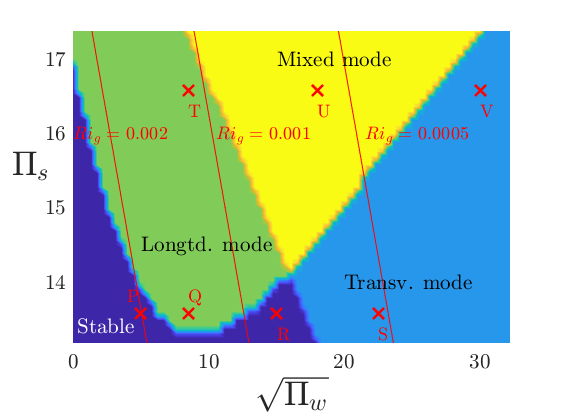}}    
\sidecaption{subfig:modes64}\raisebox{-0.9\height}{
\includegraphics[width=0.47\textwidth]{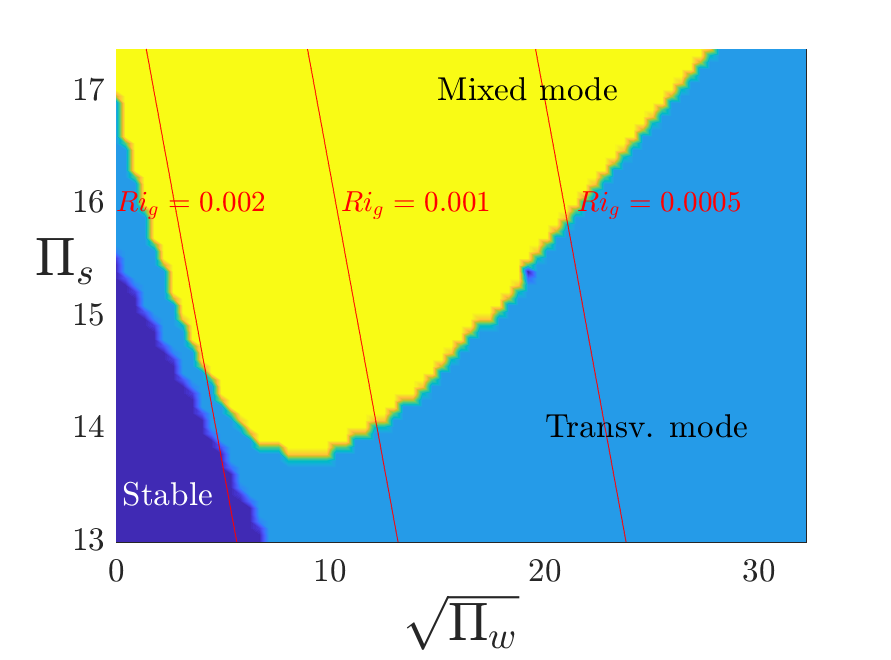}}
\caption{ Regions of different modes at slope angles $\alpha=67^{\circ}$(a) and $\alpha=61^{\circ}$(b) depending on $\Pi_s,\Pi_w$. The red lines are contours
of  $Ri_g$ number. The marked points P, Q, R, S all have  $\Pi_s=13.8$, whereas points T,U,V 
have $\Pi_s=16.5$. The transitions among these states are shown in supplementary movies obtained from direct numerical simulation of equations \ref{eqnslopemom}-\ref{eqnslopebuoy}. Movie 1: from Q(longitudinal mode) to R(stable); movie 2: from R(stable) to S(transverse mode); movie 3: from T(longitudinal mode) to U(mixed mode) and movie 4: from U(mixed mode) to V(transverse mode).}
\label{fig:modesregions}
\vspace{-3mm}
\end{figure}
For the longitudinal instability mode at the steep angle of $\alpha=67^{\circ}$, however,  an increase $\Pi_w$ only destabilizes the mode when $\Pi_w \leq 100$; beyond  the approximate value $\Pi_w\approx 100$, growing $\Pi_w$ starts to decrease the mode's growth rate, thus \textit{stabilizing} the mode, which runs counter to expectations. Since the longitudinal mode  is the dominant instability at steep angles in the absence of ambient wind, when the surface buoyancy measured by $\Pi_s$ is kept constant, increasing the ambient wind from the value corresponding to $\Pi_w=100$ onward tends to \textit{stabilize} the flow. As is known from equation \ref{eq:Ri}, the $Ri_g$ number is monotonically decreasing with respect to $\Pi_w$, so this behaviour implies that a lowering of $Ri_g$ stabilizes the base flow, which is an unexpected finding. However, since it is known that the ambient wind is monotonically destabilizing for the transverse mode, this effect can only persist until the ambient wind becomes large enough such that the growth rate of the previously dormant transverse mode overtakes that of its longitudinal counterpart, thus becoming the dominant instability. Such a complex behaviour of the stability region due to both stabilizing as well as destabilizing effects of an external flow parameter has also been discovered in \cite{schorner2016}, who reported the simultaneous stabilizing as well as destabilizing effects of topography on gravity-driven viscous film flows beyond the Nusselt regime.
\vspace{-2mm}
 \subsection{Mode transitions at steep slope angles }
 The aforementioned switching of the dominant instability from the longitudinal mode to transverse mode occurring at the steep angle of $\alpha=67^{\circ}$ is investigated here in more detail. 
 Figure \ref{fig:modesregions}a shows that the dominant instability mode is a complex function of both parameters $\Pi_w,\Pi_s$. For a fixed value of $\Pi_s<16.5$, the base flow is initially stable for $\Pi_w=0$, then becomes linearly unstable to   the longitudinal mode with increasing $\Pi_w$. When $\Pi_w$ continues to grow,depending on the value of $\Pi_s$, the flow  then becomes either stable again ($\Pi_s<14$) or susceptible to both longitudinal and transverse instability modes ($\Pi_s>14$).
 For $\Pi_w$ large enough, however, the dominant instability becomes the transverse mode.
 The effect of flow stabilization  despite lowering of $Ri_g$  and the subsequent mode switching can be observed in the  marked points P,Q,R,S,T,U,V shown in figure \ref{fig:modesregions}a. 
 These transitions predicted by linear modal analysis can also be observed in the four supplementary movies obtained from DNS data: Keeping  $\Pi_s$ constant at 13.8, movie 1 demonstrates the stabilizing transition from Q to R, whereas movie 2 shows the emergence of the transverse mode by moving from R to S; at the higher value  $\Pi_s=16.5$, movie 3 displays how the mixed mode appears by transition from T to U, whereas movie 4 indicates the weakening of the longitudinal  mode when moving from U to V.    The stabilizing effect of a flow parameter that is generally considered to be monotonically destabilizing has also been reported by \citet{gollub1980}, where an increase of the Rayleigh number was found to reduce the complexity of convective flow patterns for certain initial mean flow fields.
 
 The same contour plot for a slightly smaller angle of $\alpha=61^{\circ}$ is shown in figure  \ref{fig:modesregions}b, which indicates that the region 
 of longitudinal mode has completely vanished at this angle, in agreement with the known fact that the longitudinal instability is being dominated at smaller slope angles.
 \begin{figure}
\centering
\sidecaption{subfig:alphapi}\raisebox{-0.9\height}{
\includegraphics[width=0.445\textwidth]{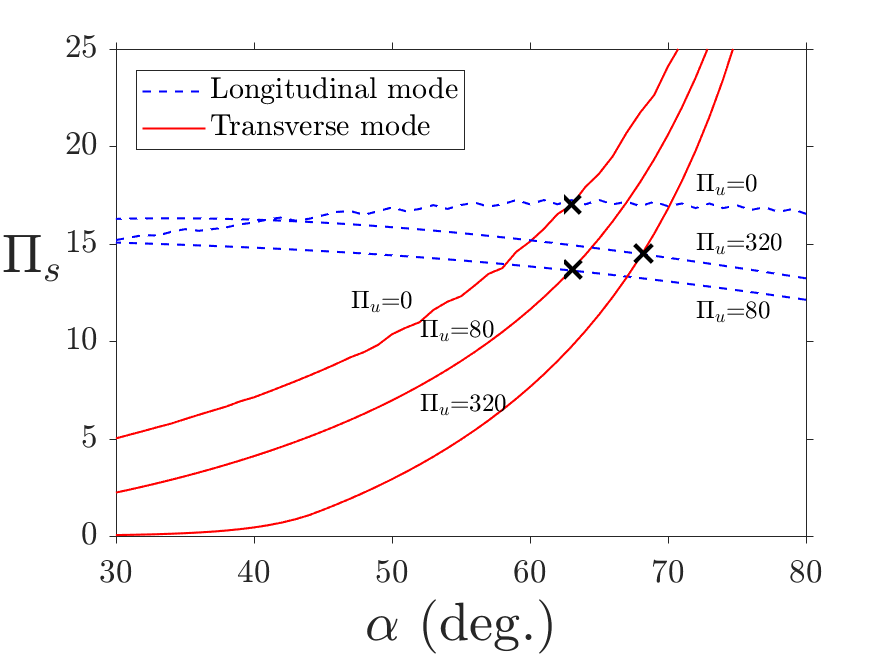}}    
\sidecaption{subfig:alphapizoom}\raisebox{-0.9\height}{
\includegraphics[width=0.445\textwidth]{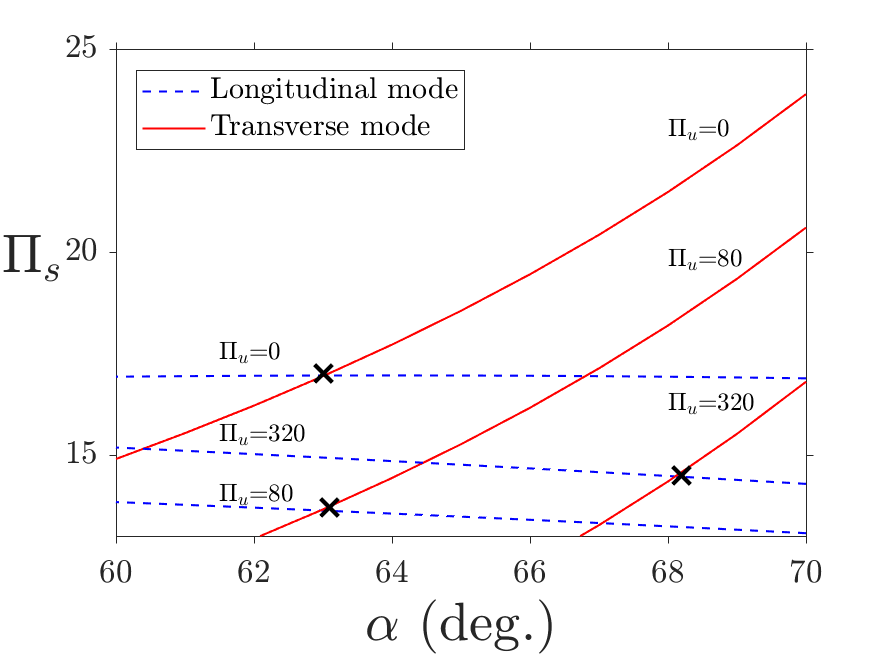}}
\caption{ \textcolor{black}{$\Pi_s-\alpha$ instability map for katabatic slope flows at $Pr=0.71$ for different ambient wind values measured by $\Pi_w$. The crosses mark the angle at which both instability modes have the same critical $\Pi_s$ threshold. Subfigure (b) zooms into the angle range in which the transition from transverse to longitudinal mode happens.  }
}\label{fig:alphaRic}
 \vspace{-3mm}
\end{figure}
 \vspace{-3mm}
\subsection{Stability at different slope angles}
As pointed out in the previous subsection, the most dangerous modes at each slope angle $\alpha$ and parameter couples $\Pi_s,\Pi_w$ are  either have pure along-slope (longitudinal mode) or pure cross-slope gradients (transverse mode). A plot of
the critical $\Pi_s$ required for the onset of each instability mode  at a specific slope angle $\alpha$ and  wind forcing number $\Pi_w$ is shown in figure \ref{fig:alphaRic}. The effect of the ambient wind on the transition slope angle $\alpha_t$ at which the dominant instability mode switches from the transverse to longitudinal mode can be clearly observed: Due to the stabilizing effect of increasing ambient wind forcing on the longitudinal mode as discussed previously, for  wind forcing number $\Pi_w$ sufficiently large, $\alpha_t$ increases beyond the value of $62^{\circ}$ found by \cite{xiao2019} in the absence of ambient wind $\Pi_w=0$ . The monotonic destabilizing   effect of growing $\Pi_w$ on the transverse mode, i.e. a lowering of its critical stability threshold over all shown angles, is also clearly visible. In particular, for  slope angles $\alpha<40^{\circ}$ and $\Pi_w=320$, we can notice that the base flow profile is unstable to the transverse mode even for  very small surface cooling as evidenced by the threshold $\Pi_s$ value close to zero.
\begin{figure}
\centering
\sidecaption{subfig:mixed37_00}\raisebox{-0.85\height}{
    \includegraphics[width=0.36\textwidth]{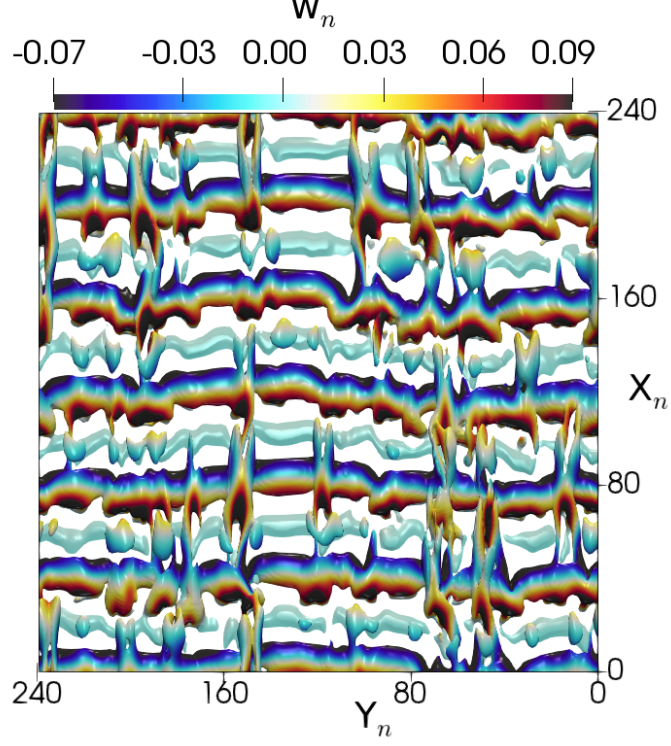}
}~
\sidecaption{subfig:mixed18_04}\raisebox{-0.85\height}{
    \includegraphics[width=0.36\textwidth]{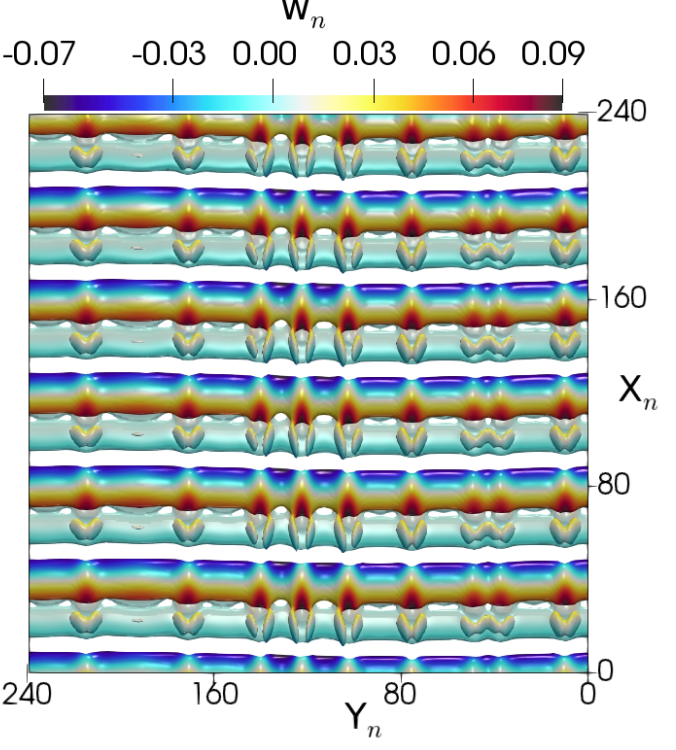}
}
\sidecaption{subfig:mixed25_00}\raisebox{-0.85\height}{
    \includegraphics[width=0.36\textwidth]{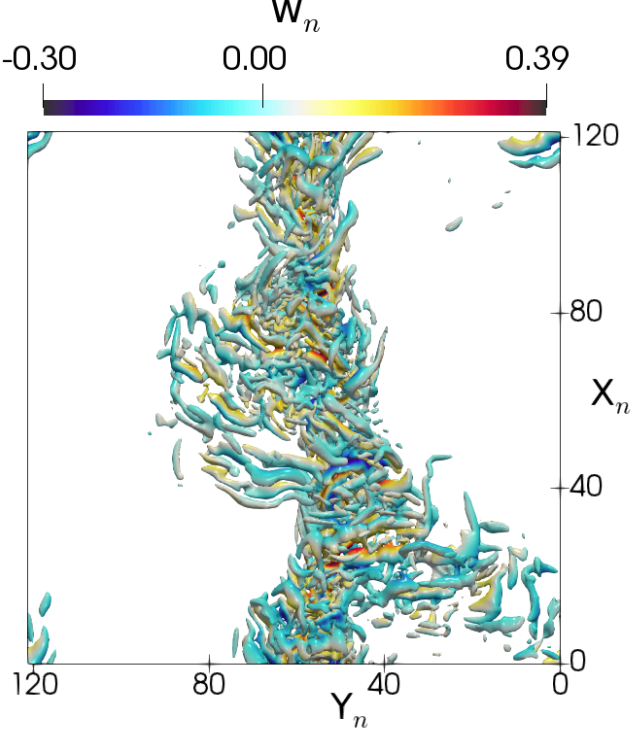}
}~
\sidecaption{subfig:mixed19_04}\raisebox{-0.85\height}{
    \includegraphics[width=0.36\textwidth]{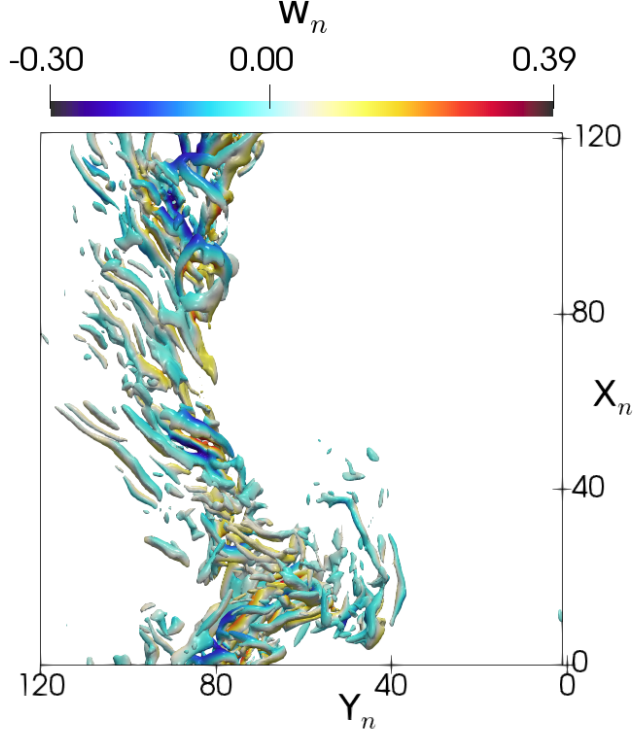}
}
\caption{Q-contour visualisations of mixed-mode instabilities. Colour represents for the slope normal velocity. Top row: $\alpha=67^{\circ}$ at constant $Ri_g \approx 5.25\times 10^{-4}$: (a)  $\Pi_s= 36.77$, $\Pi_w=0$; (b)  $\Pi_s= 18$, $\Pi_w=320$. Bottom row: $\alpha=5^{\circ}$ at constant $Ri_g=1.14\times 10^{-3}$: (c) $\Pi_s= 25$, $\Pi_w=0$; (d) $\Pi_s= 19.8$, $\Pi_w=320$. Flow is from top to bottom.}
\label{fig:mixedRi66}
\vspace{-3mm}
\end{figure}
\vspace{-4mm}
\subsection{Mixed Instability Mode}
For a steep slope angle of $\alpha=67^{\circ}$, when  $\Pi_s$ is sufficiently large,   figure~\ref{fig:alphaRic} shows that for ambient wind values corresponding to $\Pi_w=0, 320$, both the transverse and  longitudinal modes have positive growth rates. 
In order to visualise the flow field at these conditions,  the Navier-Stokes equations (\ref{eqnslopemom})-(\ref{eqnslopebuoy}) for katabatic slope flows are solved  using a Cartesian mesh, three-dimensional, bouyancy-driven incompressible flow solver \citep{Jacobsen2013}. The  settings for the direct numerical simulations are the same as adopted in \cite{xiao2019}, i.e.
 the simulation  domain is chosen to be large enough to capture multiple vortex rolls along both cross-slope and along-slope directions, and the mesh resolution ensures that there are at least two points per length scale $l_0$ in each direction.  

 To study the combined effect of the parameters $\Pi_w,\Pi_s$ on the mixed mode compared to the $Ri_g$ number, we have chosen  configurations at the same slope angle $\alpha=67^{\circ}$ and the same $Ri_g$ number, but with different combinations of $\Pi_w, \Pi_s$ determined from equation \ref{eq:Ri}.
 The first flow case contains no ambient wind and has $\Pi_s=36.77$, whereas
 the second case has a wind forcing number of $\Pi_w=320$ and a smaller $\Pi_s=18$.
 An instantaneous visualization of the results via the contour of the Q-criterion is shown in figures \ref{fig:mixedRi66}a-b, where the contour values used to obtain the plots are  the same.
 It can clearly be seen that the flow field corresponding to the larger stratification perturbation $\Pi_s=36.77$  is more unstable than its counterpart at the same $Ri_g$ number with a nonzero  wind forcing number $\Pi_w=320$.
 This serves as another confirmation of the result obtained from LSA as shown in figure \ref{fig:grcontourscrRe} where the maximal growth rate of instabilities decline along the Ri-contour when $\Pi_s$ is reduced and $\Pi_w$ is increased.
 The same comparison is made for a shallow slope  with $\alpha=5^{\circ}$,  shown in figures \ref{fig:mixedRi66}c-d.
 In the first flow configuration, we have $\Pi_s=25$ without ambient wind, whereas the second configuration has a   smaller $\Pi_s=19.8$ but a nonzero wind forcing number of $\Pi_w=320$; both flows have the same $Ri_g$ number.
 Similar like in the steep slope case, it is evident that the first flow field  with the larger stratification perturbation $\Pi_s=25$  looks more unstable and contains smaller eddies than the second flow at the same $Ri_g$ number with a nonzero wind forcing number $\Pi_w=320$. Thus, it appears that for a fixed $Ri_g$, the surface buoyancy has a stronger destabilization effect than the ambient wind higher aloft.
\vspace{-4mm}
\section{Conclusions}
We performed a linear stability analysis of the extended Prandtl model \citep{lykosov1972} for katabatic  slope flows to investigate the effect of a constant downslope ambient wind on the stability behavior of slope flows on an infinitely wide planar surface cooled from below. Our analysis has led to a new dimensionless number that we interpret as the ratio of kinetic energy of the ambient wind to the damping of kinetic energy in slope flows due to the combined action of viscosity and stable stratification. We designated this new dimensionless number the \textit{wind forcing } parameter, $\Pi_w$. We then demonstrated that the stability behavior of katabatic slope flows under the extended Prandtl model at a constant slope angle and Prandtl number is completely defined by $\Pi_w$ and the \textit{stratification perturbation} parameter ($\Pi_s$) that we have introduced earlier in \cite{xiao2019}. 
The extended Prandtl model also enables us to show analytically that the gradient Richardson number ($Ri_g$) is a  function depending on multiple parameters. $Ri_g$ is a monotonic decreasing function of $\Pi_s$ and $\Pi_w$ at a given slope angle and $Pr$ number. We conducted direct numerical simulations to further demonstrate that dynamically different slope flows do emerge under the same $Ri_g$ and the same slope angle $\alpha$. Collectively, our results show that a single $Ri_g$ criterion is ineffective to characterize the stability behaviour katabatic slope flows under the original  or extended Prandtl model. 

The types of flow instabilities that occur under the extended Prandtl model are the same as the stationary transverse mode and travelling longitudinal mode that were uncovered in \cite{xiao2019}, but their characteristics can exhibit a complex behavior as a result of  ambient wind forcing. When $\Pi_s$ is held constant, ambient wind  forcing monotonically destabilizes the stationary transverse mode. For the travelling longitudinal instability at steep slope angles, however, an increase of ambient wind forcing, within a certain range of $\Pi_w$ values, can stabilize the entire flow configuration until its value becomes sufficiently large to trigger the dormant mode of instability, which, in this case, is the stationary transverse instability. This observation runs counter to the currently held assumption that a decrease in $Ri_g$ always destabilizes the base flow. Thus, it further supports our argument that any stability criterion based solely on  $Ri_g$ number is  insufficient for slope flows under the original or the extended Prandtl model. Future subgrid-scale parameterisation schemes for stably stratified slope flows would benefit from taking into account the dependency of flow stability on the dimensionless multi-parameter space that we have laid out in the present work.

\end{document}